\documentclass[prc,nofootinbib]{revtex4}
\usepackage{amsmath,graphicx}
\usepackage{dsfont}               

\newcommand{\platzoben}{$\vphantom{{M^M}^M}$}

\begin{document}

\title{Four-quark condensates and chiral symmetry restoration in a resonance gas model}

\author{Stefan Leupold}

\affiliation{Institut f\"ur Theoretische Physik, Universit\"at
Giessen, Germany}

\begin{abstract}
As an alternative to the two-quark condensate we propose a specific four-quark 
condensate as an order parameter of chiral symmetry restoration. We show that
this four-quark condensate is closer connected to observable quantities.
Within a resonance gas model we calculate the in-medium changes of two- and
four-quark condensate as functions of temperature and baryo-chemical potential.
In this way we estimate the line of chiral symmetry restoration in the 
temperature-potential plane and also as a function of energy and baryon density.
It turns out that the line determined from the vanishing of the four-quark condensate
is extremely constant as a function of the energy density.
\end{abstract}

\maketitle

\section{Introduction}
\label{sec:intro}

In QCD, the theory of the strong interaction, the masses of the lightest quarks
are very light \cite{pdg04} as compared to typical hadronic scales. Therefore, QCD
has an approximate chiral symmetry. From the absence of degenerate chiral partners
and from the existence of very light pseudoscalar mesons one concludes that chiral
symmetry is realized in the Nambu-Goldstone phase, i.e.~it is spontaneously broken
(see e.g.~\cite{Scherer:2002tk} and references therein). Consequently, the light
pseudoscalar states are interpreted as Goldstone bosons.

It is expected that in a hot and dense strongly interacting medium chiral symmetry
gets restored. To quantify the degree of restoration of a symmetry one typically 
uses an order parameter. Indeed, concerning chiral symmetry several quantities
are conceivable as an order parameter. There are at least two considerations which 
decide how useful a possible order parameter actually might be, namely a) observability
and b) simple connection to QCD. Clearly, it would be preferable to have a 
quantity which was
observable in vacuum and also in a strongly interacting medium. In this way one could
experimentally test the process of chiral symmetry restoration. Unfortunately such a
``smoking gun'' is hard to find (cf.~e.g.~\cite{Rapp:1999ej}). 
But it should be clear that it is
appreciable as a first step, if the order parameter is at least measurable in the vacuum.
From the theoretical point of view
it would be preferable to have a quantity which fulfills criterion b), i.e.~which 
can be easily
expressed in terms of quark and gluon fields and which can be determined in lattice QCD
\cite{Creutz:1984mg} or quark model calculations. 
Basically this criterion b) boils down to the
requirement that the order parameter is a condensate.
It is the purpose of the present paper
to propose a specific four-quark condensate as a very useful order parameter.
First, however, we shall discuss order parameters which are more frequently considered.
We will especially comment how well criteria a) and b) are met.

Obviously, the appearance of Goldstone bosons is closely connected to 
spontaneous symmetry breaking. Therefore, it is natural that one can find an
order parameter of chiral symmetry restoration which is connected to the properties
of Goldstone bosons. Indeed, the pion decay constant $F_\pi$ is such a quantity.
It vanishes, if chiral symmetry is restored. Clearly, $F_\pi$ can be extracted 
experimentally from the life time of the pion. In principle, $F_\pi$ is connected to
a quark current, but the expression is non-local. In addition, it is not so easy to 
determine it from
lattice QCD (cf.~e.g.~\cite{Davies:2003ik} and references therein). 
Thus, criterion a) is met by the pion decay constant, but
criterion b) only in part.

Frequently used order
parameters are the two-quark condensates $\langle \bar u u \rangle$, 
$\langle \bar d d \rangle$
and $\langle \bar s s \rangle$ or the corresponding two- or three-flavor 
averages.\footnote{For the following qualitative discussion we will simply talk 
about {\em the} two-quark condensate without specifying the flavor. When it comes
to quantitative statements we will specify which two-quark condensate is considered.}
In lattice QCD calculations a dramatic rise of the pressure as a function of the 
temperature has been observed at a specific temperature $T_c$. This points towards
a phase transition or at least (and more likely) a rapid crossover. It has been found
that the two-quark condensate also shows a drastic change at the very same temperature
$T_c$ (more precisely, the corresponding susceptibility 
peaks at this temperature) \cite{Karsch:2001cy}. This suggests that the two-quark
condensate is a proper tool to study chiral symmetry restoration. On the other hand,
it might be that the situation is different for non-vanishing baryo-chemical 
potential, i.e.~at finite baryon density. Indeed, one can imagine scenarios where
the two-quark condensate vanishes while chiral symmetry is still 
broken (see e.g.~\cite{Birse:1996fw}). On top of that problem, criterion a) is not
met by the two-quark condensate: It is not directly connected to an observable
quantity. Here a closer look is useful:
In the standard scenario of chiral symmetry breaking ($\chi$SB) one gets \cite{gasleut1} 
(see also \cite{Knecht:1996ai})
\begin{equation}
  \label{eq:gor2}
F_\pi^2 M_\pi^2 \approx -\underbrace{(m_u+m_d)}_{=:2 m_q} \langle \bar u u \rangle 
\approx -(m_u+m_d) \langle \bar d d \rangle 
\end{equation}
(where isospin-breaking effects are neglected for the condensates). We see that it is
not simply the two-quark condensate but its combination with the quark mass
which is connected to observables\footnote{Indeed, only this combination is 
renormalization group invariant \cite{pastar}.} --- 
and also here not to one observable, but to a 
combination of two of them, the pion decay constant and the pion mass $M_\pi$.
From (\ref{eq:gor2}) it is obvious that it is possible
that the two-quark
condensate vanishes with a pion decay constant which stays more or less constant. 
A change of the pion mass would be sufficient to change the condensate.
In such a case, chiral symmetry would still be broken. Besides (\ref{eq:gor2}) there is
so far no relation derived from first principles which directly connects 
two-quark condensates with observable quantities. 

To summarize, the pion decay constant satisfies criterion a), but no so much b),
whereas it is just the other way round for the two-quark condensate.
Therefore, it is reasonable to explore also other order parameters of 
chiral symmetry breaking. Note that not much lattice QCD studies have been performed 
here. The mixed quark-gluon condensate has been studied on the lattice in 
\cite{Doi:2004jp}. We will briefly comment on this condensate 
below (section \ref{sec:goldst}). In the present paper we will mainly concentrate on
a specific four-quark condensate. In the next section we will motivate in detail why this
condensate qualifies as a useful order parameter. So far there are no lattice QCD
calculations available for the temperature dependence of this four-quark condensate.
In addition, an extension of lattice QCD to finite baryo-chemical potentials is still
complicated (cf.~e.g.~\cite{Fodor:2001pe,Allton:2002zi}). 
Therefore, we will evaluate the in-medium expectation value
of the four- and also of the two-quark condensate within a resonance gas approximation
in sections \ref{sec:resgas} and \ref{sec:goldst}. The numerical results are presented
in section \ref{sec:results}. A summary and an outlook are provided in 
section \ref{sec:sum}. In an appendix some formal issues are discussed.

\section{Generalized Weinberg sum rules}
\label{sec:genwsr}

The two main arguments which point towards spontaneous breaking of chiral symmetry
are the appearance of light states (interpreted as Goldstone bosons) and the absence
of degenerate chiral partners. In the following we focus on the second aspect.

We start with the retarded ($R$) current-current correlator \cite{Hatsuda:1993bv} 
\begin{equation}
  \label{eq:curcur}
\Pi_{\mu\nu}(q) = i \int\!\! d^4\!x \, e^{iqx} \langle R j^V_\mu(x) j^V_\nu(0) \rangle 
\end{equation}
where $j^V_\mu$ is the electromagnetic current with the quantum numbers of the 
$\rho$-meson,
\begin{equation}
  \label{eq:what-defjrho}
j^V_\mu = \frac12 \left( \bar u \gamma_\mu u - \bar d \gamma_\mu d \right) \,.  
\end{equation}
The expectation value for an arbitrary operator $\cal O$ is defined by
\begin{equation}
  \label{eq:defexpval}
\langle {\cal O} \rangle := 
\frac{{\rm Tr} \left({\cal O} e^{-\beta \, (H-\mu N)} \right) }%
{{\rm Tr} \left(e^{-\beta \, (H-\mu N)} \right) }
\end{equation}
with the Hamiltonian $H$, the baryon number operator $N$, the inverse temperature
$\beta = 1/T$, the baryo-chemical potential $\mu$
and the trace Tr is taken over all possible $n$-body states.

In a first step, we will decompose the Lorentz tensor $\Pi_{\mu\nu}(q)$. In the vacuum
we only had the structures $g_{\mu\nu}$ and $q_\mu q_\nu$ at hand.
An equilibrated medium with finite temperature and baryo-chemical potential introduces an
additional Lorentz vector $n_\mu$ which is conveniently chosen to be 
proportional to the (conserved) baryonic current and normalized to 1. In the
rest system of the medium it is simply given by $n_\mu =(1,0,0,0)$. 
Since the current (\ref{eq:what-defjrho}) is conserved by the QCD equations of motion
we have
\begin{equation}
  \label{eq:what-curconser}
q^\mu \Pi_{\mu\nu}(q) = 0 \,.
\end{equation}
We can
construct two independent projectors $L_{\mu\nu}(q)$ and $T_{\mu\nu}(q)$ which both 
still satisfy current conservation $q^\mu L_{\mu\nu}(q) = q^\mu T_{\mu\nu}(q) =0$
(cf.~e.g.~\cite{Gale:1990pn}):
\begin{subequations}
    \label{eq:what-proj}
\begin{eqnarray}
L_{\mu\nu}(q) & = & 
\frac{\left( (q \cdot n) \, q_\mu - q^2 \, n_\mu \right) 
 \left( (q \cdot n) \, q_\nu - q^2 \, n_\nu \right)}{%
q^2 \, \left( q^2 - (q \cdot n)^2 \right) }   \,,
\\[1em]
T_{\mu\nu}(q) & = & g_{\mu\nu} - \frac{q_\mu q_\nu}{q^2} - L_{\mu\nu}(q)  \,. 
\end{eqnarray}
\end{subequations}
Hence, the correlator can be decomposed in the following way:
\begin{equation}
  \Pi_{\mu\nu} = \Pi^T T_{\mu\nu} + \Pi^L L_{\mu\nu} \,.
\end{equation}
The scalar functions $\Pi^T$ and $\Pi^L$ in general depend on $q^2$ and
$(q\cdot n)^2$. In the rest frame of the medium one can equally well
state that $\Pi^{T/L}$ depends on $q^2$ and $\vec q\,^2 = (q\cdot n)^2 - q^2$
with the three-momentum $\vec q$ relative to the medium. The projectors $T_{\mu\nu}$
and $L_{\mu\nu}$ introduced in (\ref{eq:what-proj})
are chosen such that they project on states which are transverse and longitudinal,
respectively, with respect to $\vec q$.

For 
simplicity we restrict ourselves in the following to the case of vanishing $\vec q$, 
i.e.~where the current is at rest relative to the medium. 
In this case there is
no distinction between longitudinal and transverse states, i.e.~$\Pi^T$ and 
$\Pi^L$ become equal:
\begin{equation}
  \label{eq:defPi}
  \Pi^T(q^2,\vec q\,^2 =0) = \Pi^L(q^2,\vec q\,^2 =0) =: -q^2 R^V(q^2) \,. 
\end{equation}

$R^V$ has a direct physical meaning in the time-like region  $s= q^2 > 0$. 
It is related to the cross section $e^+ e^- \to $ hadrons with isospin 1 via
\cite{pastar}
\begin{equation}
  \label{eq:crosssec}
\frac{ \sigma^{I=1}(e^+ e^- \to \mbox{hadrons})}{\sigma(e^+ e^- \to \mu^+ \mu^-) }
= 12 \pi {\rm Im}R^V   \,.
\end{equation}
In addition, it is related to the decay of the $\tau$ into a neutrino and 
an even number of pions \cite{Ackerstaff:1998yj,Barate:1998uf}.

If chiral symmetry is manifestly realized, i.e.~not
spontaneously broken, the current-current correlator for the vector current
(\ref{eq:what-defjrho}) and the corresponding one for the axial-vector current 
\begin{equation}
  \label{eq:what-defjrhoA}
j^A_\mu = \frac12 \left( 
\bar u \gamma_\mu \gamma_5 u - \bar d \gamma_\mu \gamma_5 d 
\right) 
\end{equation}
must be degenerate. In other words, in this case the spectrum of hadronic states 
which couples to $j^V$ must be degenerate with the corresponding spectrum which couples
to $j^A$. These are the mentioned chiral partners. 
If chiral symmetry is spontaneously broken, there are no degenerate chiral partners
and the following Weinberg sum rules \cite{weinb} hold (in the chiral limit)
\begin{subequations}
\label{eq:wsr-wsr}
\begin{eqnarray}
\label{eq:wsr-wsr1}
\frac{1}{\pi} \int\limits_0^\infty \!\! ds \, 
\left({\rm Im}R^{V}(s) - {\rm Im}R^{A}(s) \right)
& = & F_\pi^2   \,,
\\  
\label{eq:wsr-wsr2}
\frac{1}{\pi} \int\limits_0^\infty \!\! ds \, s \,
\left({\rm Im}R^{V}(s) - {\rm Im}R^{A}(s) \right)
& = & 0   \,.
\end{eqnarray}
\end{subequations}
The quantity $R^V$ is connected to the
vector correlator according to (\ref{eq:defPi}). $R^A$ is the corresponding quantity 
for the axial-vector current (\ref{eq:what-defjrhoA}) with one subtle but important
difference: In principle, also the pion contributes to $R^A$ with a strength determined 
by the pion decay constant $F_\pi$. 
This contribution is taken out from $R^A$ and displayed
explicitly in (\ref{eq:wsr-wsr1}). Actually both quantities $R^V$ and $R^A$ can be 
observed in $\tau$ decays, if one triggers on an even (for $R^V$) or odd (for $R^A$)
number of pions in the final state \cite{Ackerstaff:1998yj,Barate:1998uf}. 

The sum rules (\ref{eq:wsr-wsr}) have been 
derived prior to the advent of QCD by using current algebra \cite{weinb}.
In other words, they are independent of QCD. In principle, one could also weight the
difference $R^V-R^A$ with higher powers of $s$. In contrast to
(\ref{eq:wsr-wsr}) such integrals are specific to QCD since quark fields and the 
strong coupling are involved (see below). 

From a practical point of view the predictions (\ref{eq:wsr-wsr}) cannot be tested
as they stand, since the integrals cover the whole range up to infinity. On the other
hand, we expect that for large $s$ the inclusive cross section in (\ref{eq:crosssec})
and therefore $R^V$ should be given by perturbative QCD \cite{pesschr}. 
Since perturbative QCD does not show chiral symmetry
breaking we expect the same behavior for the axial-vector 
correlator, i.e.~\cite{shif79,pesschr}
\begin{equation}
  \label{eq:highen}
{\rm Im}R^A(s) = {\rm Im}R^V(s) = \frac{1}{8\pi}
\left(1+\frac{\alpha_s}{\pi} \right) \qquad \mbox{for} \quad s > s_0
\end{equation}
with a threshold value $s_0$. 

Using (\ref{eq:highen}) we can restrict the integration limits in (\ref{eq:wsr-wsr})
and get
\begin{subequations}
\label{eq:s0-wsr}
\begin{eqnarray}
\label{eq:s0-wsr1}
\frac{1}{\pi} \int\limits_0^{s_0} \!\! ds \, 
\left({\rm Im}R^{V}(s) - {\rm Im}R^{A}(s) \right)
& = & F_\pi^2   \,,
\\  
\label{eq:s0-wsr2}
\frac{1}{\pi} \int\limits_0^{s_0} \!\! ds \, s \,
\left({\rm Im}R^{V}(s) - {\rm Im}R^{A}(s) \right)
& = & 0   \,,
\\
\label{eq:s0-wsr3}
\frac{1}{\pi} \int\limits_0^{s_0} \!\! ds \, s^2 \,
\left({\rm Im}R^{V}(s) - {\rm Im}R^{A}(s) \right)
& = & - \frac12 \, \pi \alpha_s
\langle {\cal O}_{\rm \chi SB} \rangle
\,.
\end{eqnarray}
\end{subequations}
We have supplemented the Weinberg sum rules by a third sum rule (\ref{eq:s0-wsr3})
which --- as already
announced --- is specific for QCD \cite{shif79}. 
It involves the four-quark condensate
\begin{eqnarray}
\langle {\cal O}_{\rm \chi SB} \rangle &= & 
\left\langle 
(\bar u \gamma_\mu \gamma_5 \lambda^a u - \bar d \gamma_\mu \gamma_5 \lambda^a d)^2 -
(\bar u \gamma_\mu  \lambda^a u - \bar d \gamma_\mu \lambda^a d)^2 
\right\rangle 
  \label{eq:wsr-fourqdef}
\end{eqnarray}
where $\lambda_a$ denotes the color Gell-Mann matrices. For an in-medium generalization
of (\ref{eq:s0-wsr}) see also \cite{Kapusta:1993hq}.

Actually the quantity $\alpha_s \langle {\cal O}_{\rm \chi SB} \rangle$ is not
a renormalization group invariant object \cite{Barfoot:1986hx}. A serious
derivation of (\ref{eq:s0-wsr3}) reveals that it must be evaluated at the 
scale $s_0$ (cf.~\cite{Dominguez:2003dr} and references therein).
There are also radiative corrections which we have not displayed explicitly.
Since the time when four-quark condensates were introduced \cite{shif79}
it is a lively discussed issue whether they can be related to the two-quark condensate
by factorization (see e.g.~\cite{Leupold:2005eq} and references therein). This discussion
is complicated by the fact that four-quark condensates are difficult to determine
experimentally. In addition, also the size of the two-quark condensate is not so well
determined: With (\ref{eq:gor2}) one can relate the two-quark condensate to the very
well known pion mass and decay constant and to the much less known current 
quark masses \cite{pdg04}. This uncertainty doubles since under the assumption of 
factorization the square of the two-quark
condensate is connected to the four-quark condensates. With all these problems in
mind we note that it has been found recently in \cite{Bordes:2005wv} that in vacuum the
four-quark condensate (\ref{eq:wsr-fourqdef}) indeed factorizes. This finding is
based on the recent data on $\tau$ decays and on a technique which we will describe
in a moment. 
On the theory side there is one particular limit in which
vacuum four-quark condensates are connected to the two-quark condensate, namely for a
large number of colors $N_c$ \cite{Novikov:1984jt}. For in-medium condensates this
is again more subtle \cite{Leupold:2005eq}. We will come
back to that point below.

From a phenomenological point of view the sum rules (\ref{eq:s0-wsr})
are still unsatisfying:
As already mentioned, nowadays experiments can address the difference of 
correlators in $\tau$ decays.
Therefore, one can explore the integrals of (\ref{eq:s0-wsr}) only up to energies
below the $\tau$ mass. It turns out that the left hand sides of the sum rules 
are not convergent as a function
of the threshold $s_0$ for values below the $\tau$ mass (squared). By proper 
combinations of (\ref{eq:s0-wsr}) one can, however, decrease the sensitivity on $s_0$
and obtain sum rules which show a much better convergence behavior already at the 
experimentally accessible energies \cite{Dominguez:2003dr,Bordes:2005wv}:
\begin{subequations}
\label{eq:WFESR-wsr}
\begin{eqnarray}
\label{eq:WFESR-wsr1}
\frac{1}{\pi s_0} \int\limits_0^{s_0} \!\! ds \, (s_0 - s)\, 
\left({\rm Im}R^{V}(s) - {\rm Im}R^{A}(s) \right)
& = & F_\pi^2   \,,
\\
\label{eq:WFESR-wsr3}
\frac{1}{\pi} \int\limits_0^{s_0} \!\! ds \, s \, (s_0 - s) \,
\left({\rm Im}R^{V}(s) - {\rm Im}R^{A}(s) \right)
& = & \frac12 \, \pi \alpha_s
\langle {\cal O}_{\rm \chi SB} \rangle
\,.
\end{eqnarray}
\end{subequations}
One can obtain the left hand sides of these sum rules from the measured spectra 
determined from $\tau$ decays. The right hand side of (\ref{eq:WFESR-wsr1}) is given 
by the experimentally measured pion decay constant. It has been found 
\cite{Bordes:2005wv} that both sides of (\ref{eq:WFESR-wsr1}) agree very well. 
In the same way, the left hand side of (\ref{eq:WFESR-wsr3}) has been determined.
Here it has been found that the result agrees well with the conjecture that the
four-quark condensate (\ref{eq:wsr-fourqdef}) factorizes into the square of the 
two-quark condensate \cite{Bordes:2005wv}. This is exactly what large-$N_c$
considerations suggest. Below we will take the finding of \cite{Bordes:2005wv} 
as a motivation to explore also the in-medium changes of the four-quark 
condensate (\ref{eq:wsr-fourqdef}) in the large-$N_c$ approximation. Note that we do not
claim here that an {\em arbitrary} four-quark condensate is {\em always} connected to 
the square of the two-quark condensate. First of all, as we will see below, even for the 
four-quark condensate at hand, its in-medium change is {\em not} given by the change
of the two-quark condensate (see also \cite{Leupold:2005eq}). Second, from the point
of view of chiral symmetry breaking, it makes sense to connect a chirally odd
four-quark condensate, e.g.~(\ref{eq:wsr-fourqdef}), to the two-quark condensate.
Concerning chirally invariant four-quark condensates\footnote{which appear e.g.~in 
QCD sum rules for the $\omega$ meson \cite{Steinmueller:2005}}, on the other hand, 
it is not so clear whether there should be any deeper connection to the two-quark
condensate. In the present work we concentrate on one specific four-quark condensate,
namely (\ref{eq:wsr-fourqdef}). In vacuum, this condensate seems to 
factorize \cite{Bordes:2005wv}.

From (\ref{eq:WFESR-wsr}) it is obvious that both quantities $F_\pi^2$ and the 
four-quark condensate (\ref{eq:wsr-fourqdef})
can be regarded as order parameters of chiral symmetry breaking ($\chi$SB). They must
vanish if chiral symmetry is restored. To be precise, in a medium the relevant 
definition of $F_\pi$ is the coupling of the pion to the {\em temporal} component
of the axial-vector current (\ref{eq:what-defjrhoA}) \cite{Meissner:2001gz}.
Concerning their role as order parameters, we want to stress again that
$F_\pi^2$ and $\alpha_s \langle {\cal O}_{\rm \chi SB} \rangle$ are actually more 
advantageous as compared
to the two-quark condensate  $\langle \bar q q \rangle$ which is the ``standard'' 
order parameter: The former are connected to quantities which are in principle
measurable, namely the left hand sides of relations (\ref{eq:WFESR-wsr}).\footnote{In
practice this is a very tedious task: Access on $R^V$ can be obtained from dilepton
spectra as discussed above; see also 
relation (\ref{eq:crosssec}). Indeed, one tries to measure $R^V$ also in a 
medium (see e.g.~\cite{Rapp:1999ej}).
In vacuum, $R^A$ is related to $\tau$ decays \cite{Ackerstaff:1998yj,Barate:1998uf}
as already discussed. This is obviously difficult to measure in a medium.}

The in-medium changes of $F_\pi^2$ are addressed in \cite{Gasser:1986vb,Gasser:1987zq} 
for a non-interacting pion gas (density $\rho_\pi$)
and e.g.~in \cite{Meissner:2001gz} for a 
non-interacting (and cold) Fermi sphere of nucleons (density $\rho_N$). 
For finite temperatures (pion gas) one gets
\begin{equation}
  \label{eq:fpitempdep}
F_\pi^2(T) = F^2_\pi \, \left(1- \frac{4 \rho_\pi}{ 3 F_\pi^2} \right)
\end{equation}
with the (scalar) pion density
\begin{equation}
  \label{eq:wsr-piondens}
\rho_\pi = 3 \int \! \frac{d^3 k}{ (2 \pi)^3 \, 2 E_k} \, \frac{1}{ e^{E_k/T} - 1} \;
\stackrel{M_\pi \to \, 0}{\longrightarrow} \; \frac18 \, T^2
\end{equation}
and the pion energy $E_k = \sqrt{\vec k^2 + M_\pi^2}$.
For cold nuclear matter (nucleon Fermi sphere) 
the change of $F_\pi^2$ can be related to some
low-energy constants of the chiral pion-nucleon Lagrangian. For details we refer to
\cite{Meissner:2001gz}. Numerically the result is 
\begin{equation}
  \label{eq:fpidensdep}
F^2_\pi(\rho_N) = F^2_\pi \, \left(1-\frac{\rho_N}{ \rho_0} (0.52 \pm 0.08) \right)
\end{equation}
where $\rho_0$ denotes nuclear saturation density. These results can be contrasted with
the corresponding changes of the 
two-quark condensate $\langle \bar u u + \bar d d \rangle$ 
\cite{Gasser:1986vb,Gerber:1989tt}:
\begin{eqnarray}
\frac{\langle \bar u u + \bar d d \rangle_{{\rm pionic\; med.}}}{%
\langle \bar u u  + \bar d d \rangle_{{\rm vac}}} 
& \approx & 
1 - \frac{ \rho_\pi}{F_\pi^2}
  \label{eq:fintemp2q}
\end{eqnarray}
and \cite{Drukarev:1991fs}
\begin{equation}
\label{eq:dropfindens}
\frac{\langle \bar u u + \bar d d \rangle_{{\rm nucl.\; med.}}}{%
\langle \bar u u  + \bar d d \rangle_{{\rm vac}}} 
 \approx 
1- \frac{\rho_N \sigma_N}{F_\pi^2 m_\pi^2} \approx 1 - \frac13 \frac{\rho_N}{ \rho_0} \,.
\end{equation}
The nucleon sigma term 
\begin{equation}
  \label{eq:defnuclsig}
\sigma_N =   m_q \, \frac{d m_N}{d m_q} \approx 45\,{\rm MeV}
\end{equation}
which appears in (\ref{eq:dropfindens}) can be obtained within the framework of
chiral perturbation theory from low-energy $\pi N$ scattering \cite{Gasser:1991ce}. 
In principle, there are corrections to these linear-density results 
(\ref{eq:fpitempdep}), (\ref{eq:fpidensdep}), (\ref{eq:fintemp2q}) and
(\ref{eq:dropfindens}) which are of higher powers in the respective density. They
involve correlations between the constituents of the medium.
For the quark condensate in a pion gas this has been worked out in \cite{Gerber:1989tt}.
For all other cases model independent predictions are hard to make for the terms 
beyond linear order in the density.

Comparing (\ref{eq:fpitempdep}) with (\ref{eq:fintemp2q}) or (\ref{eq:fpidensdep})
with (\ref{eq:dropfindens}) shows that a bold extrapolation of these formulae to the
point where the respective order parameter vanishes would lead to different critical
densities. This indicates a breakdown of the linear-density approximation for at least
one of the compared respective quantities, probably for both.

Beyond these special cases of a (non-interacting) pion or nucleon gas
the changes of $F_\pi^2$ are hard to estimate (except for specific models). The
situation is somewhat better for the four-quark condensate (\ref{eq:wsr-fourqdef})
to which we turn for the rest of this work. We will compare the obtained results
to the in-medium changes of the two-quark 
condensate $\langle \bar u u + \bar d d \rangle$. We note in passing that the
in-medium behavior of $\langle \bar s s \rangle$ is different from the one of
$\langle \bar u u + \bar d d \rangle$ \cite{Pelaez:2002xf}. Since we started out from
the non-strange correlators $R^V$ and $R^A$, a comparison to 
$\langle \bar u u + \bar d d \rangle$ is most appropriate.

\section{Resonance gas model}
\label{sec:resgas}

The in-medium changes described above in equations (\ref{eq:fpitempdep}), 
(\ref{eq:fpidensdep}), (\ref{eq:fintemp2q}) and (\ref{eq:dropfindens}) have been
determined under the assumption that the respective medium is described by a 
non-interacting gas of the respective most abundant particles (pions for a medium with
finite temperature and nucleons for cold nuclear matter). Of course, to describe
a medium with finite temperature and finite baryo-chemical potential one can
easily generalize the previous results by using a gas of pions and nucleons 
(and anti-nucleons).\footnote{For our qualitative discussion we restrict ourselves to
flavor $SU(2)$. For our quantitative analysis we will include strangeness.}
Clearly, this is not the full story --- except for low particle densities. Indeed,
the in-medium expectation value (\ref{eq:defexpval}) 
is defined with respect to all possible states. Beyond the
consideration of single-particle states one has to take into account $n$-particle
states with $n>1$, e.g.~two-pion, pion-nucleon, nucleon-nucleon, three-pion states
and so on. The influence of $n$-particle states is accompanied by higher powers
in the respective densities and therefore suppressed for low temperatures and low
baryo-chemical potentials. This justifies the previous results as the respective leading
terms in a low-density expansion. However, we want to understand now how to go
beyond this linear-density approximation --- at least approximately.

One might wonder why we only talked about pions and nucleons so far and not about
their excitations. Indeed we will come to the excitations soon. From a principal
point of view, however, we stress that there is no need to include them in the traces
which appear in (\ref{eq:defexpval}): For a complete set of states it is sufficient
to consider all single- and many-body states built from the {\em stable} states
(for flavor $SU(2)$: pions, nucleons and anti-nucleons). 
As demonstrated long time ago in 
\cite{Dashen:1969} the (unstable\footnote{with respect to the strong interaction}) 
excitations emerge from the scattering phase shifts which come into play when 
considering many-body states in the traces in (\ref{eq:defexpval}). E.g.~a hadron 
resonance which is formed in pion-nucleon scattering is automatically taken into
account when one considers the contributions of two-particle states (pion-nucleon)
in the traces in (\ref{eq:defexpval}). 

On the other hand, it is in practice intractable to calculate the contributions
of all $n$-body states to the expectation value (\ref{eq:defexpval}). Now we can turn
the previous argument around:
Physically
we expect that the contribution of a many-body state is the more important, the
larger the many-body correlation is. Clearly, a large correlation is found, if a
hadronic resonance is formed. Therefore one might approximate the complete set of
(stable) many-body states by a sum over all 
one-body resonance states \cite{Hagedorn:1965st}.
Below we will evaluate the in-medium two- and four-quark condensate in this resonance
gas approximation
(see also e.g.~\cite{Gerber:1989tt,Karsch:2003vd,Karsch:2003zq} for 
successful applications of that idea). 

The following aspects should have become clear from the previous discussion:\\
a) The resonance gas approximation takes into account part of the many-body
correlations of the stable states which form the medium. In that way one goes
beyond the linear-density approximation discussed above.\\
b) The resonance gas approximation is not a systematic expansion in powers of the
densities of (stable) medium constituents. One merely tries to take into account
the most important parts of the higher density terms.\\
c) The resonance gas approximation can only make sense, if the many-body correlations
are governed by hadron resonances. This {\em excludes} the application to systems
with low temperatures and large baryo-chemical potential. In such systems 
nucleon-nucleon correlations are the most important ones. 

In the resonance gas approximation we find
\begin{equation}
  \label{eq:lindensintro}
\langle {\cal O} \rangle_{\rm med.} \approx \langle 0 \vert {\cal O} \vert 0 \rangle
+ \sum\limits_X \rho_X \langle X \vert {\cal O} \vert X \rangle
\end{equation}
where $\vert 0 \rangle$ denotes the vacuum state and ${\cal O}$ an arbitrary operator. 
$\rho_X$ is the scalar density of states $X$:
\begin{equation}
  \label{eq:densXdef}
\rho_X = \int \frac{d^3 k}{(2\pi)^3} \, \frac{m_X}{E_X} \, n_{F/B}(E_X-s\mu)
\end{equation}
with 
\begin{equation}
  \label{eq:distrFBdef}
n_{F/B}(E) = 
\left[\exp\left( \frac{E}{T} \right) \pm 1 \right]^{-1} \,,
\end{equation}
$E_X = \sqrt{m_X^2+{\vec k}^2}$ and $s=1$ for baryons, $s=-1$ for antibaryons
and $s=0$ for mesons. The $\pm$ sign and the label $F/B$ refer to baryons and mesons, 
respectively. We sum over all states $X$ identified by the 
particle data group \cite{pdg04}.

The normalization of the state $\vert X \rangle$ is chosen to be
\begin{equation}
  \label{eq:normX}
\langle X(\vec k) \vert X(\vec k') \rangle = 
\frac{E_X}{m_X} \, (2\pi)^3 \delta(\vec k - \vec k')  \,.
\end{equation}
We already note here that this normalization is appropriate for heavy states
since it has a proper non-relativistic limit. On the other hand, it is not quite
adequate for Goldstone bosons
where we would like to have states with a properly defined chiral limit. Of course,
one can choose a different normalization, if one changes the definition of
the density (\ref{eq:densXdef}) accordingly. We will
come back to that point below. 

We have to discuss the evaluation of (\ref{eq:lindensintro}) separately for
the two- and four-quark condensates. Concerning the type of hadron $X$ we also will 
distinguish in the following between Goldstone bosons and other hadrons. If $X$
is a Goldstone boson, the expectation values 
$\langle X \vert \bar u u + \bar d d \vert X \rangle$ and 
$\langle X \vert {\cal O}_{\rm \chi SB} \vert X \rangle$ can be calculated using
current algebra \cite{Eletsky:1992xd,Hatsuda:1993bv}. We postpone this discussion
to section \ref{sec:goldst}. 

To calculate the contribution of non-Goldstone states to the in-medium part of the 
two-quark condensate we follow the approach of \cite{Gerber:1989tt} and generalize 
it to finite baryo-chemical potential (see also \cite{Toublan:2004ks,Tawfik:2005qh}). We 
approximate\footnote{Note that this relation becomes so simple using the normalization
(\ref{eq:normX}).}
\begin{equation}
  \label{eq:twoquarkapprox}
\langle X \vert \bar u u + \bar d d \vert X \rangle \approx
\langle X \vert u^\dagger u + d^\dagger d \vert X \rangle = 
\left\{
  \begin{array}{ll}
3 - N_s & \mbox{for baryons,} \\
2 - N_s & \mbox{for mesons,} 
  \end{array}
\right. 
\end{equation}
where $N_s$ denotes the number of strange quarks in the state $X$. For the 
sigma term \cite{Gasser:1991ce} this amounts to the approximation
\begin{equation}
  \label{eq:sigmaterm}
\sigma_X = \left\{
  \begin{array}{ll}
(3 - N_s)\, m_q & \mbox{for baryons,} \\
(2 - N_s)\, m_q & \mbox{for mesons.} 
  \end{array}
\right. 
\end{equation}
For the nucleon and the $\Delta(1232)$ 
we can get an idea how good this approximation is: Using
$m_q \approx 7\,$MeV and (\ref{eq:defnuclsig})
we find that (\ref{eq:sigmaterm}) underestimates $\sigma_N$ by about a factor
of 2. On the other hand, a recent determination of $\sigma_\Delta$ yields 20.6 MeV
\cite{Bernard:2005fy} in good agreement with (\ref{eq:sigmaterm}). 
We will explore the
uncertainty induced by the estimate (\ref{eq:twoquarkapprox}) by an additional
calculation where we simply increase $\sigma_X$ by a factor of 2 for each particle
species. The in-medium two-quark condensate (but not the four-quark condensate)
has also been calculated within a resonance
gas approximation in \cite{Toublan:2004ks}. There an estimate 
different from (\ref{eq:twoquarkapprox}) has been used for the quark
condensate within a resonance. We will come back to this point below.

Next we have to evaluate the expectation values appearing in (\ref{eq:lindensintro}) 
for the four-quark condensate (\ref{eq:wsr-fourqdef}). For the Goldstone bosons we note
again that this can be performed using current algebra as will be discussed below 
in section \ref{sec:goldst}. For all other hadrons it is difficult to get an estimate
for $\langle X \vert {\cal O}_{\rm \chi SB} \vert X \rangle$. As already noted after
equation (\ref{eq:wsr-fourqdef}) it is even a problem to pin down the vacuum
expectation values of four-quark operators. 
For $\langle 0 \vert {\cal O}_{\rm \chi SB} \vert 0 \rangle$ it has been shown in
\cite{Bordes:2005wv} that it factorizes in the vacuum. In line with that finding, we use
in the following the large-$N_c$ expansion 
\cite{'tHooft:1974jz,witten} where $N_c$ denotes the number of colors. 
For the operator of interest, ${\cal O}_{\rm \chi SB}$, we 
obtain \cite{Leupold:2004gh,Leupold:2005eq}
\begin{equation}
  \label{eq:scalingvac}
\langle 0 \vert {\cal O}_{\rm \chi SB} \vert 0 \rangle = O(N_c^2) = 
8 \, \langle 0 \vert \bar u u \vert 0 \rangle^2 + o(N_c) \,;
\end{equation}
\begin{equation}
  \label{eq:scalingbar}
\langle X \vert {\cal O}_{\rm \chi SB} \vert X \rangle = O(N_c^2) = 
8 \, \langle 0 \vert \bar u u \vert 0 \rangle 
\langle X \vert \bar u u + \bar d d \vert X \rangle  + o(N_c)  \,,
\end{equation}
if $X$ is a baryon;
\begin{equation}
  \label{eq:scalingmes}
\langle X \vert {\cal O}_{\rm \chi SB} \vert X \rangle = o(N_c) \,,
\end{equation}
if $X$ is a meson.

Therefore, it seems that to leading order in $1/N_c$ we only must consider baryonic 
states and
we can relate the in-medium four-quark condensate to two-quark condensates.
However, there is an additional implicit $N_c$-dependence in $\rho_X$
which enters (\ref{eq:lindensintro}). We have
shifted the discussion of that issue to appendix \ref{sec:app1}. 
The outcome of this discussion
is that one should keep the respective leading order in
$1/N_c$ for every type of hadron to get a serious estimate for all regions of
$T$ and $\mu$ (as long as $T$ is not too small, cf.~point c) in the discussion above).

To obtain the respective leading $1/N_c$ contribution for a specific type of hadron
we need a closer look at the derivation of (\ref{eq:scalingbar}) and 
(\ref{eq:scalingmes}). We will summarize the essential points of \cite{Leupold:2005eq}
and apply it to our case of interest. First we note that the proof for 
(\ref{eq:scalingbar}) and (\ref{eq:scalingmes}) can be given for four-quark 
condensates involving white (color singlet) quark-antiquark operators. 
On the other hand, our condensate of interest (\ref{eq:wsr-fourqdef})
involves color octet operators. However, we can use a Fierz transformation to
obtain
\begin{eqnarray}
\label{eq:fierz}
\lefteqn{\langle 
(\bar u \gamma_\mu \gamma_5 {\lambda_a} u 
- \bar d \gamma_\mu \gamma_5 {\lambda_a} d)^2 -
(\bar u \gamma_\mu  {\lambda_a} u 
- \bar d \gamma_\mu {\lambda_a} d)^2 
\rangle = } \nonumber \\[1.5em] &&
2 \, \langle (\underbrace{\bar u u + \bar d d}_{{ \sim f_0}})^2 \rangle 
+ 2 \, \langle (\underbrace{\bar u i \gamma_5 u 
                        - \bar d i \gamma_5 d}_{{ \sim \pi^0}})^2 \rangle 
- 8 \, \langle \underbrace{\bar u i \gamma_5 d}_{{ \sim \pi^-}} \, 
                    \bar d i \gamma_5 u \rangle
+ 2 \, \langle (\underbrace{\bar u i \gamma_5 u 
                         + \bar d i \gamma_5 d}_{{ \sim \eta, \eta'}})^2 \rangle 
+ 2 \, \langle (\underbrace{\bar u u - \bar d d}_{{ \sim a_0^0}})^2 \rangle 
- 8 \, \langle \underbrace{\bar u d}_{{ \sim a_0^-}} \, \bar d u \rangle
\nonumber \\[1em]  && {}
- \frac{2}{N_c} \, \langle 
(\underbrace{\bar u \gamma_\mu \gamma_5  u 
             - \bar d \gamma_\mu \gamma_5  d}_{{ \sim \pi^0,a_1^0}})^2 -
(\underbrace{\bar u \gamma_\mu  u - \bar d \gamma_\mu  d}_{{ \sim \rho^0}})^2 
\rangle  \,.
\end{eqnarray}
For an easier discussion and for orientation we have attributed meson states to the 
respective operators which have the
same quantum numbers. The four-quark operators which appear on the
right hand side of (\ref{eq:fierz}) all have the generic 
structure $\bar q \Gamma q \, \bar q \Gamma' q$ where $\Gamma$ and $\Gamma'$ denote
spinor and flavor, but no color structure. It has been shown in \cite{Leupold:2005eq}
that for this type of four-quark operators
the following holds for expectation values with respect to single-particle states $X$:
\begin{eqnarray}
  \label{eq:fact-non}
\langle X \vert \bar q \Gamma q \, \bar q \Gamma' q \vert X \rangle & = &
\langle X \vert \bar q \Gamma q \vert X \rangle \, 
\langle 0 \vert \bar q \Gamma' q \vert 0 \rangle +
\langle 0 \vert \bar q \Gamma q \vert 0 \rangle \, 
\langle X \vert  \bar q \Gamma' q \vert X \rangle 
\nonumber \\ && {}
+ \langle X \vert \bar q \Gamma q \vert 0 \rangle \, 
\langle 0 \vert \bar q \Gamma' q  \vert X \rangle
+ \langle 0 \vert \bar q \Gamma q \vert X \rangle \, 
\langle X \vert \bar q \Gamma' q  \vert 0 \rangle
+ \mbox{terms subleading in $1/N_c$.}
\end{eqnarray}
Obviously, the first two terms on the right hand side of (\ref{eq:fact-non}) can 
contribute for every type of $X$, but only if $\Gamma$ and $\Gamma'$ have the quantum
numbers of the vacuum. This is only fulfilled by the first term appearing on the right
hand side of (\ref{eq:fierz}). We can use again (\ref{eq:twoquarkapprox}) to evaluate
the respective contribution. This is the factorized part of the four-quark condensate.

Next we turn to the third and fourth term given on the
right hand side of (\ref{eq:fact-non}). These terms are contributions beyond
factorization.
First we note that these terms do 
not exist, if $X$ is a baryon: A quark-antiquark operator $\bar q \Gamma q$ cannot
create a baryon $X$ from the vacuum. This simple fact finally 
leads to (\ref{eq:scalingbar}).
If $X$ is a meson, the third and fourth term given on the
right hand side of (\ref{eq:fact-non}) only exist, if the
quantum numbers of the meson match to the ones of $\bar q \Gamma q$ and
$\bar q \Gamma' q$. If this is the case, we need to know the overlap of the
meson $X$ with the quark currents $\bar q \Gamma q$ and $\bar q \Gamma' q$. In principle,
such a question can be addressed within lattice QCD. However, for higher excited
hadrons this is a difficult task. In principle, we would need this information.
In practice, however, we will find that we can safely neglect the respective 
contributions. To see how this comes about, let us first discuss which hadrons are of 
relevance here:  
In view of (\ref{eq:fierz}) we see that particles which yield contributions in leading
$1/N_c$ order on top of 
the factorized part have quantum numbers
of a) Goldstone bosons, b) $f_0$ or c) $a_0$. We disregard the vector and axial-vector
mesons here since their contributions are $1/N_c$ suppressed in (\ref{eq:fierz}).
Concerning the group a) we note that
only the excited states are of concern here: Goldstone bosons will be fully taken 
into account in section \ref{sec:goldst}. There, no approximations, especially no
large-$N_c$ expansion is used besides the current algebra technique. 
Nonetheless, we note that we have checked explicitly that the current algebra
calculations are not in contradiction to the large-$N_c$ considerations used here.
Still we have to be concerned with excitations of the Goldstone bosons and with
$f_0$ and $a_0$ states. All these states have masses of about 1 GeV or 
higher \cite{pdg04}.\footnote{We disregard the $f_0(400-1200)$ which we consider as a 
loosely bound meson-meson molecule and not a genuine hadron resonance
state \cite{Colangelo:2001df,Pelaez:2003dy}.}
As we will show below (cf.~figure \ref{fig:comp}) 
these higher excited meson states 
are irrelevant from a practical point of view: For temperatures below, say, 200 MeV
relevant contributions to the
sum (\ref{eq:lindensintro}) over all resonances come from the low-lying mesons
(pseudoscalar and vector nonet) and from baryons. Higher lying mesons are
strongly Boltzmann suppressed and --- in contrast to high lying baryons --- do not have 
large degeneracy factors. It turns out that the sum of all baryon resonances ---
albeit also Boltzmann suppressed --- adds up to a non-negligible contribution whereas
the sum of all higher lying mesons remains irrelevant. Therefore, in practice we do not
need the third and fourth term on the right hand side of (\ref{eq:fact-non})
--- except for Goldstone bosons to which we turn in the next section. Finally we
note that in the spirit of the large-$N_c$ expansion we also
treat the $\eta'$ as a Goldstone boson (cf.~e.g.~\cite{Borasoy:2004ua}
and references therein).

\section{Goldstone bosons}
\label{sec:goldst}

For Goldstone bosons we can use current algebra to calculate the contribution
$\langle X \vert {\cal O} \vert X \rangle$ 
in (\ref{eq:lindensintro}). First, however, we shall choose a normalization
which is more appropriate for Goldstone bosons. Instead of (\ref{eq:densXdef})
and (\ref{eq:normX}) we use for Goldstone bosons \cite{Eletsky:1992xd}
\begin{equation}
  \label{eq:densGdef}
\rho_G = \int \frac{d^3 k}{(2\pi)^3} \, \frac{1}{2 E_G} \, n_{B}(E_G)
\end{equation}
and
\begin{equation}
  \label{eq:normG}
\langle G(\vec k) \vert G(\vec k') \rangle = 
 2 E_G \, (2\pi)^3 \delta(\vec k - \vec k')  \,.
\end{equation}
With this normalization (which has a well-defined chiral limit) we get using
current algebra e.g.~for pions \cite{Eletsky:1992xd,Hatsuda:1993bv}:
\begin{equation}
  \label{eq:pcac}
\langle \pi^a \vert {\cal O} \vert \pi^b \rangle =
- \frac{1}{F_\pi^2} \, \langle 0 \vert [Q_5^a,[Q_5^b,{\cal O}]] \vert 0 \rangle
\end{equation}
with
\begin{equation}
  \label{eq:defQ5}
  Q_5^a = \frac12 \int \!d^3x \, \bar q(x) \gamma_0 \gamma_5 \tau^a q(x)
\end{equation}
and a Gell-Mann flavor matrix $\tau^a$. 

Obviously this can be generalized to other
Goldstone bosons as well. As already mentioned we treat the whole pseudoscalar
nonet as Goldstone bosons in the spirit of the large-$N_c$ approximation. We use the
decay constant corresponding to the type of particle in the 
denominator of (\ref{eq:pcac}). For the flavor singlet we use 
$\tau^0 = \sqrt{2/3} \, \mathds{1}$. 

It turns out that $\eta$ and $\eta'$ do not
contribute at all to $\langle {\cal O}_{\rm \chi SB} \rangle_{\rm med.}$.
They do contribute to the two-quark condensate 
(see (\ref{eq:drop4q}) and (\ref{eq:drop2q}) below). 
Why there is no contribution to the four-quark condensate can be explicitly
seen, if one does not use current algebra directly for (\ref{eq:wsr-fourqdef}), but
applies it separately to all the terms appearing on the right hand side 
of (\ref{eq:fierz}), i.e.~after Fierz transformation. If $X$ is an $\eta$ or $\eta'$,
the first operator on the right hand side of (\ref{eq:fierz}) contributes via
the first two terms on the right hand side of (\ref{eq:fact-non}). The fourth
operator on the right hand side of (\ref{eq:fierz}) contributes via the third
and fourth term on the right hand side of (\ref{eq:fact-non}).
We have checked explicitly that in total these contributions indeed cancel.

With relations like (\ref{eq:pcac}) all expectation values can be traced back 
to vacuum expectation
values. For the four-quark operator ${\cal O}_{\rm \chi SB}$ this leads to 
similar four-quark operators which in part deviate by their flavor content. We assume
flavor symmetry of the vacuum to relate all resulting vacuum four-quark condensates
to $\langle 0 \vert {\cal O}_{\rm \chi SB} \vert 0 \rangle$. {\em No} factorization
(\ref{eq:scalingvac}) is needed here. 

With the same technique (\ref{eq:pcac}) the two-quark 
condensate $\langle \bar u u + \bar d d \rangle_{\rm med.}$ is related to the vacuum
two-quark condensates $\langle 0 \vert \bar u u + \bar d d \vert 0 \rangle$ and
$\langle 0 \vert \bar s s \vert 0 \rangle$
(the latter appears using kaons instead of pions in (\ref{eq:pcac})). 
Again we assume flavor symmetry of the
vacuum: $\langle 0 \vert \bar s s \vert 0 \rangle \approx 
\langle 0 \vert \bar u u + \bar d d \vert 0 \rangle /2$.

Finally we want to comment on another order parameter of chiral symmetry breaking,
the mixed quark-gluon condensate\footnote{For the following comment we do not distinguish
between flavor $SU(2)$ and flavor $SU(3)$.} 
$\langle \bar q \sigma_{\mu\nu} G^{\mu\nu} q \rangle$
with the gluon field strength $G^{\mu\nu}$. In \cite{Doi:2004jp} it has been
found that in lattice QCD the temperature dependence of the two-quark and the mixed
condensate are practically the same. It has been suggested in \cite{Doi:2004jp}
that this finding points towards a universal behavior of the order parameters. 
With the techniques presented here we can 
evaluate $\langle \bar q \sigma_{\mu\nu} G^{\mu\nu} q \rangle_{\rm med.}$ for a gas of
Goldstone bosons. It turns out that the operators $Q_5^a$ given in (\ref{eq:defQ5})
do not distinguish between $\mathds{1}$, which appears in the two-quark condensate, 
and $\sigma_{\mu\nu} G^{\mu\nu}$, which appears in the mixed condensate. Therefore,
for a gas of Goldstone bosons one gets
\begin{equation}
  \label{eq:mix2q}
\frac{\langle \bar q \sigma_{\mu\nu} G^{\mu\nu} q \rangle_{\rm med.}}%
{\langle \bar q q \rangle_{\rm med.}} = 
\frac{\langle 0 \vert \bar q \sigma_{\mu\nu} G^{\mu\nu} q \vert 0 \rangle}%
{\langle 0 \vert \bar q q \vert 0 \rangle} \,.
\end{equation}
On the other hand, for the four-quark condensate (\ref{eq:wsr-fourqdef})
the temperature dependence is different (see below). 
Therefore, it is questionable to conclude only
from the temperature dependence of the two-quark and the mixed condensate that
there is a universal behavior. Note that we do not claim that a gas of Goldstone bosons
is equivalent to a lattice QCD calculation. We just have presented an example where
the temperature dependence of two-quark and mixed condensate agrees, but where 
other order parameters show a different behavior.

\section{results}
\label{sec:results}

Now we can take all terms together. We find for the four-quark condensate:
\begin{eqnarray}
  \label{eq:drop4q}
\frac{\langle {\cal O}_{\rm \chi SB} \rangle_{\rm med.}}%
{\langle 0 \vert {\cal O}_{\rm \chi SB} \vert 0 \rangle} & = & 
1 
- \frac{2}{\pi^2 F_\pi^2} \int\limits_0^\infty \!\! dk \, \frac{k^2}{E_\pi} \,
n_B(E_\pi)
- \frac{1}{\pi^2 F_K^2} \int\limits_0^\infty \!\! dk \, \frac{k^2}{E_K} \,
n_B(E_K)
\nonumber \\ && {}
- \frac{m_q}{\pi^2 F_\pi^2 M_\pi^2} \, \sum\limits_B \, (3-N_s) \, u \, m_B 
\int\limits_0^\infty \!\! dk \, \frac{k^2}{E_B} \, n_F(E_B - s \mu)
\nonumber \\ && {}
- \frac{m_q}{\pi^2 F_\pi^2 M_\pi^2} \, \sum\limits_M \, (2-N_s) \, u \, m_M 
\int\limits_0^\infty \!\! dk \, \frac{k^2}{E_M} \, n_B(E_M)
\end{eqnarray}
where $M$ denotes meson resonances except for $\pi$, $K$, $\eta$ and $\eta'$.
The factor $u$ parameterizes the uncertainty connected to the estimate
(\ref{eq:twoquarkapprox}). In the figures below we will show calculations using
$u=1$ in agreement with (\ref{eq:twoquarkapprox}). In addition, we will use $u=2$
to explore the uncertainty of our estimate (cf.~the 
discussion after (\ref{eq:sigmaterm})).

For the two-quark condensate we get
\begin{eqnarray}
  \label{eq:drop2q}
\frac{\langle \bar u u + \bar d d \rangle_{\rm med.}}%
{\langle 0 \vert \bar u u + \bar d d \vert 0 \rangle} & = & 
1 
- \frac{3}{4\pi^2 F_\pi^2} \int\limits_0^\infty \!\! dk \, \frac{k^2}{E_\pi} \,
n_B(E_\pi)
- \frac{1}{2\pi^2 F_K^2} \int\limits_0^\infty \!\! dk \, \frac{k^2}{E_K} \,
n_B(E_K)
\nonumber \\ && {}
- \frac{1}{12\pi^2 F_\eta^2} \int\limits_0^\infty \!\! dk \, \frac{k^2}{E_\eta} \,
n_B(E_\eta)
- \frac{1}{6\pi^2 F_{\eta'}^2} \int\limits_0^\infty \!\! dk \, \frac{k^2}{E_{\eta'}} \,
n_B(E_{\eta'})
\nonumber \\ && {}
- \frac{m_q}{2\pi^2 F_\pi^2 M_\pi^2} \, \sum\limits_B \, (3-N_s) \, u \, m_B 
\int\limits_0^\infty \!\! dk \, \frac{k^2}{E_B} \, n_F(E_B - s \mu)
\nonumber \\ && {}
- \frac{m_q}{2\pi^2 F_\pi^2 M_\pi^2} \, \sum\limits_M \, (2-N_s) \, u \, m_M 
\int\limits_0^\infty \!\! dk \, \frac{k^2}{E_M} \, n_B(E_M)  \,.
\end{eqnarray}
For the numerics we use the parameter values listed in table \ref{tab:numval}.
\begin{table}[thb]
\centering
\begin{tabular}{|c|c|c|}
\hline
quantity & value in MeV & ref. \\ \hline \hline
$F_\pi$ \platzoben  & 92.4 & \cite{pdg04}  \\ \hline
$M_\pi$ \platzoben  & 140. & \cite{pdg04}  \\ \hline
$m_q$ \platzoben  & 7. & \cite{pdg04}  \\ \hline
$F_K$ \platzoben  & 113. & \cite{pdg04}  \\ \hline
$F_\eta$ \platzoben  & 124. & \cite{Borasoy:2004ua}  \\ \hline
$F_{\eta'}$ \platzoben  & 107. & \cite{Borasoy:2004ua}  \\ \hline
\end{tabular}
\caption{Parameter values used for actual calculations.}
\label{tab:numval}
\end{table}

As already noted we have treated the kaons as Goldstone bosons. Of course, one can
study which formula we would have obtained, if we treated the kaon in the same way as the
non-Goldstone bosons. This yields an opportunity to check our approximations.
If kaons were considered as standard mesons instead of Goldstone bosons we observe first
of all that there is indeed a relative factor 2 between the two kaon contributions
in (\ref{eq:drop4q}) and (\ref{eq:drop2q}) --- this is what one would expect
from factorization: In (\ref{eq:fact-non}) the first two terms on the right hand side
give the leading $1/N_c$ contribution, if $X$ is a kaon. The third and fourth term
drop out since the quantum numbers of the operators appearing in (\ref{eq:fierz})
do not match to the kaon. Of course, we see this same relative factor 2 in all 
non-Goldstone contributions in (\ref{eq:drop2q}) compared to (\ref{eq:drop4q}).
Next we compare the kaon contribution in (\ref{eq:drop2q})
to the last contribution in (\ref{eq:drop2q}): Instead of $1/(2\pi^2 F_K^2)$ we would 
have for $u=1$ (the factor 4 comes
from the four types of kaons)
\begin{equation}
  \label{eq:compK}
4 \, \frac{m_q}{2\pi^2 F_\pi^2 M_\pi^2} \, M_K \,.
\end{equation}
Indeed these two numbers deviate only by 6\%. This finding gives us some confidence
that our approximations are not too bad.

\begin{figure}[htbp]
\includegraphics[keepaspectratio,width=0.8\textwidth]{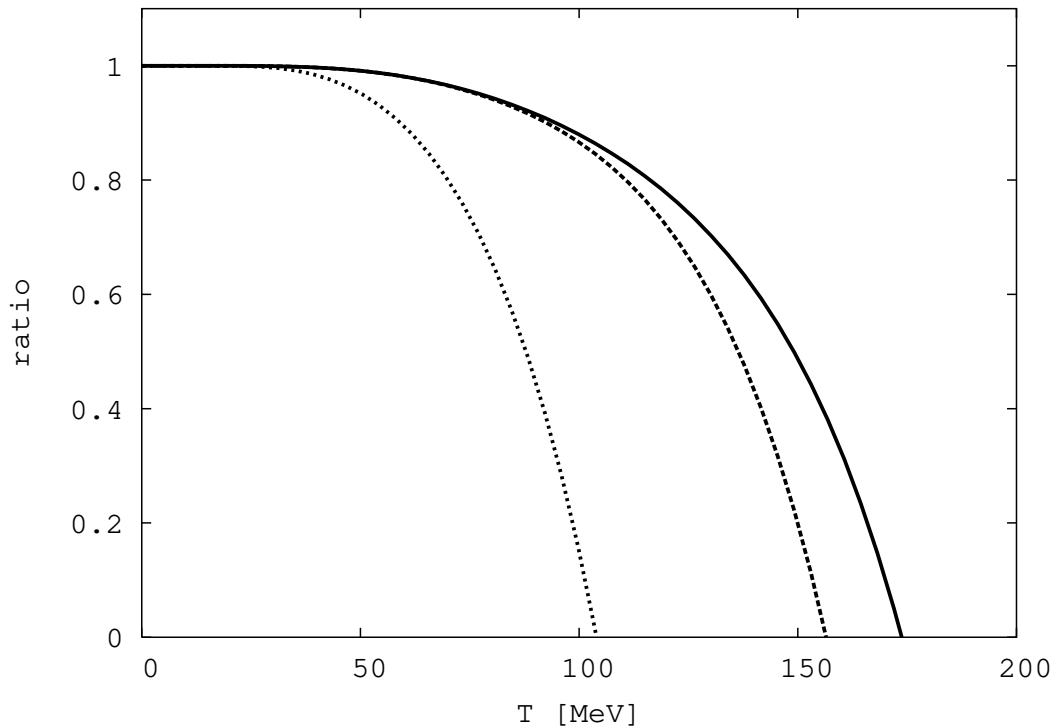}
\caption{Drop of the four-quark condensate \eqref{eq:drop4q} (for $u=1$)
as a function of the 
temperature $T$ for three different baryo-chemical potentials: $\mu =0$ (full line),
$\mu = 400\,$MeV (dashed), $\mu = 800\,$MeV (dotted).} 
\label{fig:3d}  
\end{figure}
\begin{figure}
\includegraphics[keepaspectratio,width=0.8\textwidth]{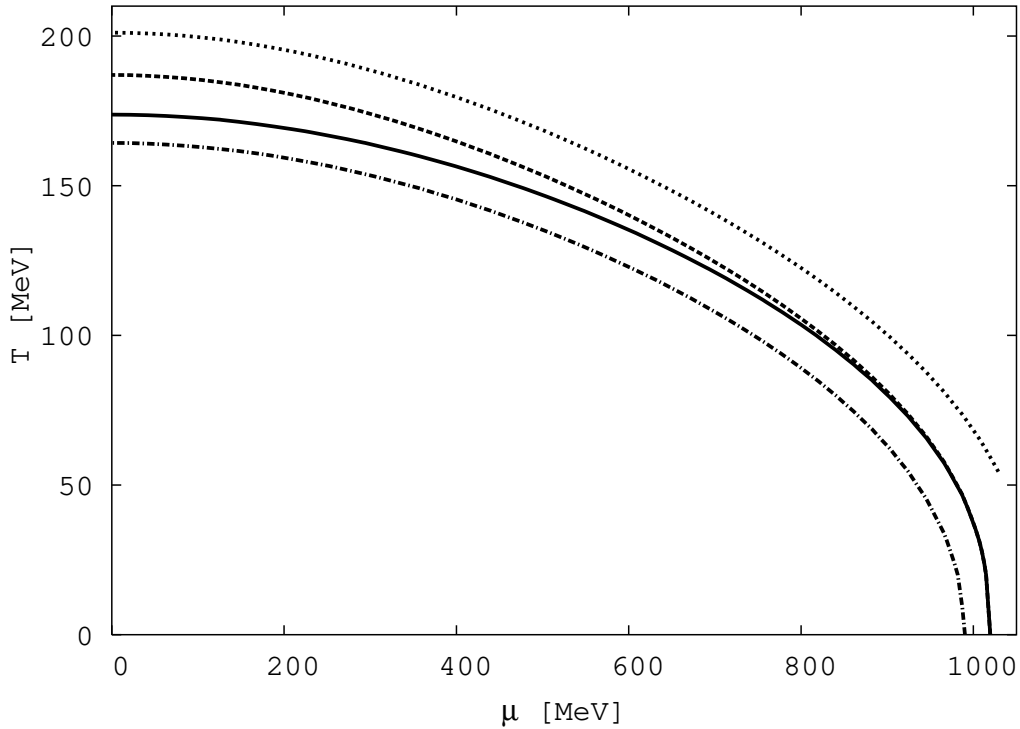}
\caption{Vanishing of two- and four-quark condensates in temperature-potential plane
for different values of $u$. Lines enumerated from top to bottom.
Line 1: Vanishing of two-quark condensate (\ref{eq:drop2q}) with $u=1$.
Line 2: Vanishing of two-quark condensate with $u=2$.
Line 3: Vanishing of four-quark condensate (\ref{eq:drop4q}) with $u=1$.
Line 4: Vanishing of four-quark condensate with $u=2$.}
\label{fig:transline}  
\end{figure}

With (\ref{eq:drop4q}) at hand we can study how the four-quark condensate changes
in a medium. This is depicted in figure \ref{fig:3d}. We see that the four-quark
condensate drops both with temperature $T$ and baryo-chemical potential $\mu$. 
Of course, for the two-quark
condensate there emerges a plot qualitatively similar to the one shown 
in figure \ref{fig:3d}. We will not display these results, but concentrate in the
following on the line where the respective condensate vanishes. This is depicted
in figure \ref{fig:transline}. 
From inspecting (\ref{eq:drop4q}) and (\ref{eq:drop2q}) one can already deduce
that the four-quark condensate drops faster than the two-quark condensate and that
the drop increases with the uncertainty parameter $u$. We observe these qualitative
features also in figure \ref{fig:transline}. Quantitatively we deduce first of all
that for vanishing chemical potential all lines yield a reasonable transition 
temperature between about 165 and 200 MeV. Without a bias which order parameter
might be preferable we can view the lines in figure \ref{fig:transline} as estimates
for the transition line to a chirally restored state. The highest and the lowest
line mark the uncertainty of the estimates. On the other hand, as discussed in great
detail in section \ref{sec:genwsr} we prefer the use of the four-quark condensate
as a quantity which is closer connected to observables. As we will see next, the
transition line extracted from the vanishing of the four-quark condensate shows
an additional interesting feature.

An appealing aspect of the resonance gas approximation is the fact that one can easily
calculate the energy and baryon density which correspond to given temperature
and baryo-chemical potential:
\begin{eqnarray}
  \label{eq:endens}
\varepsilon & = & \sum\limits_B \int \! \frac{d^3 k}{(2\pi)^3} \, E_B \, n_F(E_B-s\mu) 
+ \sum\limits_M \int \! \frac{d^3 k}{(2\pi)^3} \, E_M \, n_B(E_M) 
\nonumber \\ & = &
\frac{1}{2\pi^2} \sum\limits_{B, \, {\rm no\,}\bar B} \,
\int\limits_0^\infty \!\! dk \, k^2 E_B \, 
\left[ n_F(E_B-\mu) + n_F(E_B+\mu) \right] 
+ \frac{1}{2\pi^2} \sum\limits_{M}
\int\limits_0^\infty \!\! dk \, k^2 E_M \, n_B(E_M) 
\end{eqnarray}
and
\begin{eqnarray}
  \label{eq:bardens}
\rho & = & 
\frac{1}{2\pi^2} \sum\limits_{B, \, {\rm no\,}\bar B} \,
\int\limits_0^\infty \!\! dk \, k^2 \, 
\left[ n_F(E_B-\mu) - n_F(E_B+\mu) \right] \,.
\end{eqnarray}
Note that here $M$ denotes all mesons, including the Goldstone bosons. 

\begin{figure}
\includegraphics[keepaspectratio,width=0.8\textwidth]{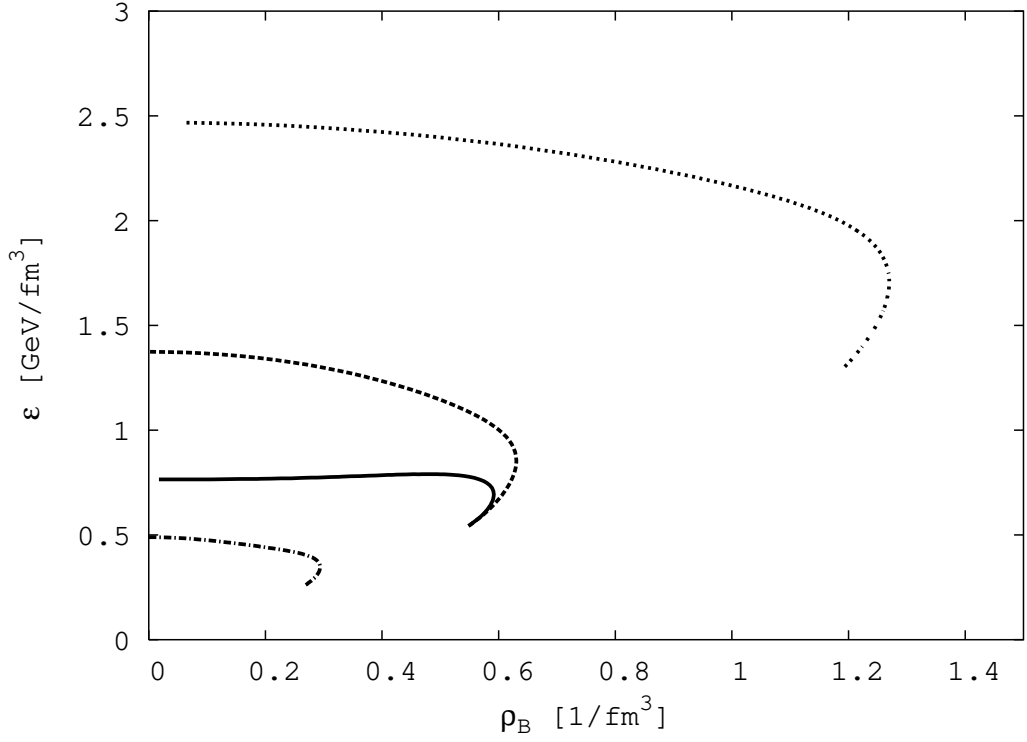}
\caption{Vanishing of two- and four-quark condensates 
in the plane of energy and baryon density for different values of $u$.
Same line code as in figure \ref{fig:transline}.} 
\label{fig:transline-dens}  
\end{figure}
Now we can study the vanishing of the condensates also as functions of the energy and
baryon density. Figure \ref{fig:transline-dens} shows the corresponding results.
We observe first of all that the critical energy density widely differs for the
four different lines. This is not surprising: Differences in the temperature
multiply to a high power for the determination of the corresponding energy density
(for massless states we would have $\epsilon \sim T^4$). Figure \ref{fig:transline-dens}
reveals a particularly interesting
aspect: The energy density is extremely constant along the line where the four-quark
condensate vanishes. Only for large baryon densities a different behavior sets in.
This is the region of low temperatures (numerically $T < 100$ MeV)
where we distrust the resonance gas approximation anyway. 
The critical
energy density obtained from the vanishing of the four-quark condensate for $u=1$
is about 0.8 GeV/fm$^3$. This agrees very well with lattice 
QCD results \cite{Karsch:2003vd}.  
Note that the critical energy density obtained from the four-quark
condensate for $u=2$ is not quite that constant --- albeit still more stable than
the critical lines obtained from the two-quark condensate. We recall that $u=1$
corresponds to our genuine estimate (\ref{eq:twoquarkapprox}) whereas $u=2$ has
been introduced by hand to get an idea about the uncertainties.
We conclude that the four-quark condensate seems to be a very useful order parameter
to determine the line of chiral symmetry restoration --- at least within the
resonance gas approximation.

\begin{figure}
\includegraphics[keepaspectratio,width=0.8\textwidth]{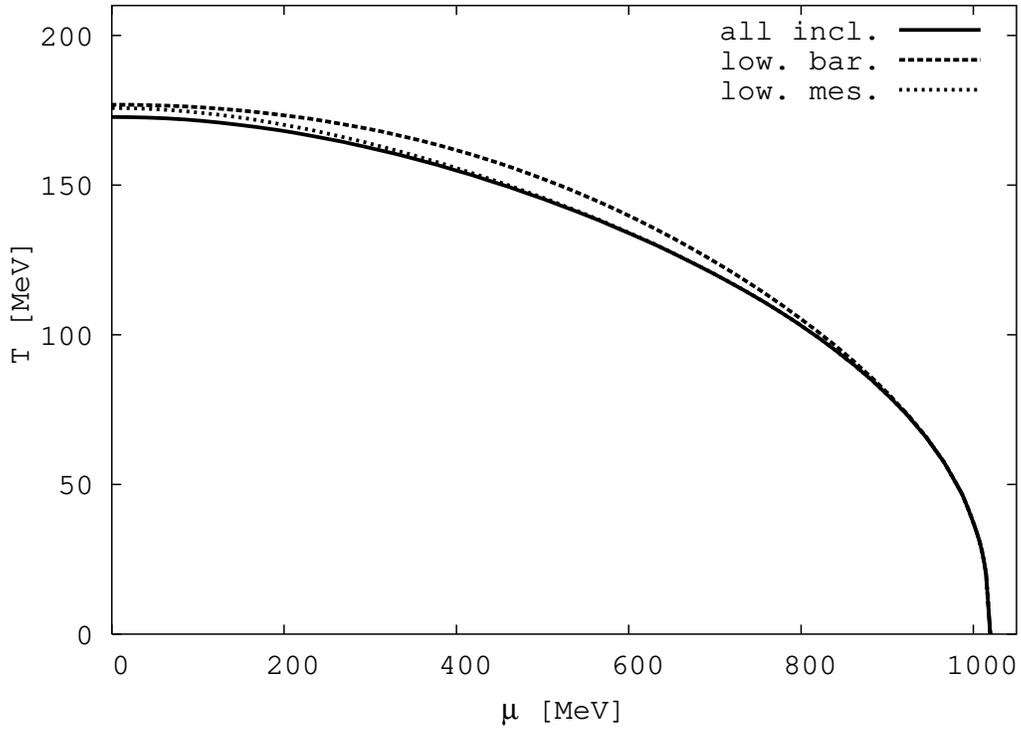}
\caption{Vanishing of the four-quark condensate (for $u=1$) 
for different sets of resonances.
Full line: all mesons, all baryons; dotted line: lowest two meson nonets, all baryons;
dashed line: all mesons, lowest two baryon multiplets.} 
\label{fig:comp}  
\end{figure}

Finally we shall discuss the uncertainties induced by the fact that not all
resonances listed by the particle data group \cite{pdg04}
are well established: We have checked that it makes no difference, whether we use
all baryons of \cite{pdg04} or only the 3- and 4-star baryons. Figure \ref{fig:comp}
illustrates that it also makes no difference,
whether one uses all mesons or only the lowest pseudoscalar and vector meson nonets.
Using only the lowest octet and decuplet baryons instead of all baryons makes a 
difference, especially in an intermediate range of chemical potentials
(see figure \ref{fig:comp}). We conclude that
high-mass baryons are to some extent important due to their large multiplicities 
whereas high-mass mesons
are not relevant. We recall that this is a fortunate situation, since we do not know the
overlap of high-mass mesons with the corresponding quark currents which appears 
in \eqref{eq:fact-non}.

\section{Summary and outlook}
\label{sec:sum}

The present work consists of two parts: First, a motivation why the four-quark condensate
(\ref{eq:wsr-fourqdef}) should be regarded as a useful order parameter of chiral
symmetry restoration. Second, a calculation within the resonance gas approximation
to determine how this four-quark condensate drops as a function of temperature and 
baryo-chemical potential and in particular where it vanishes. As our favorite prediction
for the line of chiral symmetry restoration
we regard the full lines in figures \ref{fig:transline} and \ref{fig:transline-dens}.
This yields a temperature $T_c \approx 174\,$MeV at vanishing chemical potential
and an energy density $\epsilon_c \approx 0.8\,$GeV/fm$^3$ which stays remarkably
constant as a function of baryon density or baryo-chemical potential.

The corresponding line which one obtains from the vanishing of the two-quark instead
of the four-quark condensate does not show the same stability for the energy density
$\epsilon_c$. This seems to be in contrast to the work \cite{Toublan:2004ks}
where also the in-medium change of the two-quark condensate has been determined
within a resonance gas approximation. Here, we have to point out that our estimate
for the resonance sigma terms (\ref{eq:sigmaterm}) adopted from \cite{Gerber:1989tt}
differs from the estimate utilized in \cite{Toublan:2004ks}.
In the latter work the approximation \cite{Tawfik:2005qh} 
\begin{equation}
  \label{eq:sigmaalt}
\sigma_X \approx M_\pi^2 \, \frac{A}{m_X}
\end{equation}
with $0.9 \le A \le 1.2$ has been used for all resonances, 
no matter whether these are baryons or mesons. Such
an approach does not agree with large-$N_c$ counting rules \cite{Leupold:2005eq}: 
For baryons the sigma term should scale
with $N_c$, whereas for mesons it should be constant. This requirement is satisfied
by our approach: The factor $3-N_s$ in (\ref{eq:twoquarkapprox}) and
(\ref{eq:sigmaterm}) is actually $N_c - N_s$ since a baryon consists of $N_c$
quarks \cite{witten}. In contrast, in the approximation (\ref{eq:sigmaalt}) the
sigma terms of baryons scale with $1/N_c$ instead of $N_c$ since baryon masses ---
which appear in the denominator of (\ref{eq:sigmaalt}) --- scale with 
$N_c$. Note that the numerator in (\ref{eq:sigmaalt}) cannot scale with $N_c$.
Otherwise the scaling law for the meson resonances would be violated.
Due to that deficiency we prefer our estimate (\ref{eq:sigmaterm}).

\begin{figure}
\includegraphics[keepaspectratio,width=0.8\textwidth]{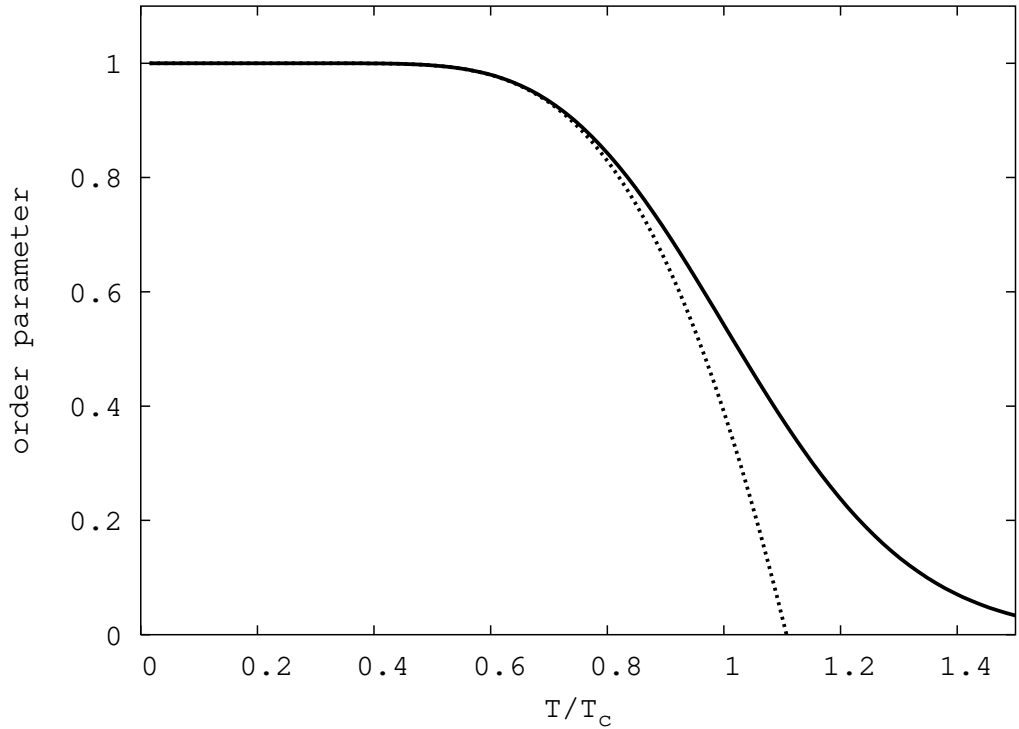}
\caption{Sketch of the real behavior of an order parameter as a function of
temperature (full line) and of a reasonable approximation obtained from the 
low-density degrees of freedom (dotted line).} 
\label{fig:true}  
\end{figure}

Next we would like to discuss our approach in a more general context: In principle,
we do not expect that any of our order parameters completely vanishes for high
densities/temperatures. Only if chiral symmetry was an {\em exact} symmetry of QCD,
the order parameters would exactly vanish in the chirally restored phase. In
reality with finite current quark masses we expect a sizable drop of the order 
parameters at the phase transition or crossover point. After this transition point
the order parameter presumably levels off. We have sketched the expected
behavior in figure \ref{fig:true}.
Clearly the exact behavior
of the order parameters near the transition point is hard to extract from
low-density expansions or more general from an approach which uses only the
degrees of freedom which are relevant below the transition point. Therefore we should
not expect to obtain a completely accurate description of the transition region.
Nonetheless, if one has a model which still works reasonably well in the region
where the drastic change of the order parameter sets in, the estimated transition
point should not deviate too much from the real one. Such a scenario is sketched
in figure \ref{fig:true}.

For the evaluation of the in-medium behavior of the four-quark condensate we have used
two crucial approximations: First, a resonance gas to model
the in-medium system, and second, large-$N_c$ arguments to evaluate the four-quark
condensate for the non-Goldstone bosons. Concerning the latter approximation it is
unfortunately hard to check its validity for the real world of $N_c=3$. For general
arguments in favor of the large-$N_c$ approximation we refer 
to \cite{'tHooft:1974jz,witten} and to the large amount of works which cite these
seminal papers. 

Concerning the resonance gas approximation there is at least one aspect which might
lead to additional non-negligible contributions beyond the ones obtained from the 
resonance gas: Since the pions are very light, the threshold for their interactions 
($2 M_\pi$) is not significantly Boltzmann suppressed for the temperatures we are 
interested in (close to 200 MeV). On the other hand, if all the pion-pion interactions 
were mediated by
{\em narrow} resonances, the relevant scale would not be set by the threshold but by
the resonance masses. Indeed, there are arguments in favor of a resonance saturation
of the low-energy constants which determine the 
pion-pion interaction \cite{gasleut1,eckgas,Ecker:1989yg}. However, these resonances,
in particular the $\rho$-meson, are not narrow. Therefore a significant interaction
already shows up at energies below the $\rho$-meson mass. Hence, pion-pion interactions
should be taken into account beyond the resonance gas approximation. For the
two-quark condensate at finite temperature and vanishing baryo-chemical potential
these interactions have been evaluated up to three loops within chiral perturbation
theory in \cite{Gerber:1989tt}. It turned out that the effects induced by these 
interactions are not completely negligible, but small as compared to the combined
effects from the non-interacting pions, nucleons and the resonances \cite{Gerber:1989tt}.
To clarify whether this also is true for the four-quark condensate is devoted to
future work. Another possible and straightforward extension of our approach is the 
inclusion of
isospin and strangeness chemical potential. Also this is beyond the
scope of the present work.

\acknowledgments S.L.~acknowledges the support of the European 
Community-Research Infrastructure
Activity under the FP6 ``Structuring the European Research Area'' programme
(HadronPhysics, contract number RII3-CT-2004-506078).

\appendix

\section{Resonance density and large number of colors}
\label{sec:app1}

From \eqref{eq:scalingbar} and \eqref{eq:scalingmes} it seems that to leading order 
in $1/N_c$ we only must consider baryonic states.
However, there is an additional implicit $N_c$-dependence in $\rho_X$ as we will
discuss now. We will find that we have to distinguish different regimes for
the baryo-chemical potential. This is intimately connected to the fact that the
masses of baryons scale with $N_c$ \cite{witten}:
\begin{equation}
  \label{eq:barmass}
m_X = O(N_c) \,,
\end{equation}
if $X$ is a baryon, whereas
\begin{equation}
  \label{eq:mesmass}
m_X = O(N_c^0) \,,
\end{equation}
if $X$ is a meson. The reason is that a baryon consists of $N_c$ quarks while
a meson is a quark-antiquark state.

We start with the case that the chemical potential
is small compared to all baryon masses. For temperatures which are reasonable
for hadronic degrees of freedom we have for baryons
\begin{equation}
  \label{eq:smallmu}
\sqrt{m_X^2+{\vec k}^2} - \mu > m_X - \mu \gg T  \,.
\end{equation}
In this case we can replace the Fermi distribution in (\ref{eq:densXdef}) by a
Boltzmann distribution and we can treat the problem non-relativistically. This yields
for baryons
\begin{eqnarray}
  \label{eq:densXboltz}
\rho_X & \approx & \int \frac{d^3 k}{(2\pi)^3} \, \frac{m_X}{E_k} \, 
\exp\left( -\frac{E_k-\mu}{T} \right) \approx
\exp\left( -\frac{m_X-\mu}{T} \right) 
\int \frac{d^3 k}{(2\pi)^3} \, \exp\left( - \frac{{\vec k}^2}{ 2 m_X T} \right)
\sim  \exp\left( -\frac{m_X-\mu}{T} \right) (m_X T)^{3/2} 
\nonumber \\ 
& = & o\left(N_c^{3/2} \exp(-N_c) \right)  \,.
\end{eqnarray}
For mesons there is no $N_c$-dependence,
$\rho_X = O(N_c^0)$.
Since the expectation value of the four-quark condensate is enhanced by $N_c$ for
baryons (cf.~(\ref{eq:scalingbar}) and (\ref{eq:scalingmes})), 
we find in total that the relative importance of a baryonic contribution
as compared to a mesonic one is $o(N_c^{5/2} \exp(-N_c))$. 
In the large-$N_c$ limit an exponential suppression overwhelms any power enhancement.
Therefore, for small baryo-chemical potentials the in-medium change of the four-quark
condensate is dominated by mesons. 

The situation changes, however, for larger baryo-chemical potentials to which we turn
next:
For $m_X-\mu = O(N_c^0)$ there is no implicit $N_c$-dependence left in $\rho_X$, i.e.
$\rho_X = O(N_c^0)$ even for baryons.\footnote{We note in passing that the 
contributions of antibaryons are always, i.e.~for arbitrary
positive baryo-chemical potential, suppressed by $N_c^{5/2} \exp(-N_c)$.}
In this case, the in-medium four-quark condensate is dominated by baryons. 

To get a serious estimate for all regions we should keep the respective leading order in
$1/N_c$ for every type of hadron.

\bibliography{literature}
\bibliographystyle{apsrev}

\end{document}